\documentclass[twoside,12pt]{article}
\usepackage[utf8]{inputenc}
\usepackage{extsizes}
\usepackage[super,sort&compress,comma]{natbib} 
\usepackage[version=3]{mhchem}
\usepackage[left=3cm, right=3cm, top=1.785cm, bottom=2.0cm]{geometry}
\usepackage{balance}
\usepackage{mathptmx}
\usepackage{tcolorbox}
\usepackage{etoolbox}
\usepackage{sectsty}
\usepackage{graphicx} 
\usepackage{lastpage}
\usepackage[format=plain,justification=justified,singlelinecheck=false,font={stretch=1.125,small,sf},labelfont=bf,labelsep=space]{caption}
\usepackage{float}
\usepackage{fancyhdr}
\usepackage{fnpos}
\usepackage[english]{babel}
\addto{\captionsenglish}{%
  \renewcommand{\refname}{Notes and references}
}
\usepackage{array}
\usepackage{droidsans}
\usepackage{amssymb}
\usepackage{charter}
\usepackage[T1]{fontenc}
\usepackage[usenames,dvipsnames]{xcolor}
\usepackage{setspace}
\usepackage[compact]{titlesec}
\usepackage{hyperref}
\usepackage{siunitx}
\usepackage{booktabs}

\usepackage{epstopdf}
\definecolor{cream}{RGB}{222,217,201}

\titleformat{\section}          
  {\bfseries\fontsize{14pt}{16pt}\selectfont} 
  {\thesection}                
  {1em}                        
  {}                           

  \titleformat{\subsection}          
  {\bfseries\fontsize{12pt}{14pt}\selectfont} 
  {\thesubsection}                
  {1em}                        
  {}                           
\begin{document}

\begin{flushleft}
{\fontsize{18pt}{20pt}\selectfont\textbf{Discovering new photovoltaics using optimal transport theory}}
\end{flushleft}
{\fontsize{12pt}{14pt}\selectfont Matthew A. H. Walker\textsuperscript{a}, Zibo Zhou\textsuperscript{c}, Junayd Ul Islam\textsuperscript{d}, and Keith T. Butler\textsuperscript{b,*}}\\\\


Searching by chemical and structural analogy is one of the most commonly used and successful approaches to materials discovery. However, formulating this task for algorithmic implementation raises the question of how we define similar materials. Methods have been proposed for searching materials space using vectors based on chemical composition and functional fragments in the material. Descriptors for structural similarity have also been proposed. However, the question of how to incorporate and balance structural and compositional similarity measures in a single metric remains open. Here, we adapt methods developed for calculating distances between undirected graphs and apply them to crystalline materials similarity. The Fused Gromov-Wasserstein (FGW) metric uses optimal transport theory to map between two graphs considering a balance of the graph structure  and the information present at the nodes of the graph (atoms in crystals). We apply the method to exploring new photovoltaic materials. We demonstrate that FGW is competitive with embeddings from an equivariant graph neural network, trained on $> 10^6$ materials, despite minimal training. We then apply FGW to a discovery campaign to identify materials from the Materials Project database that have not previously been explored as photovoltaics, but have similarities to known high-efficiency materials. After validating predictions with hybrid density functional theory, we identify seven previously unexplored high-efficiency photovoltaic absorber candidates, including \ce{Cs_5Sb_8}, which is found to have a predicted SLME of $> 30\%$ and to be thermodynamically stable. The FGW approach demonstrates the power of strong inductive biases for developing metrics for materials exploration with minimal training data.
\vspace{1cm}
\section*{Introduction}
The need for efficient and sustainable photovoltaic (PV) materials to meet the requirement for renewable energy sources to mitigate the worst of climate change continues to grow\cite{haegel2019terawatt}. Photovoltaic discovery has historically involved considerable trial and error based on similarity approaches: using knowledge of chemistry through, for instance, trends of the periodic table, to move from Se to Si to GaAs, CdTe, and other III-V heterojunctions\cite{goodman1958prediction, pamplin1964systematic}. More recently, after the hybrid halide perovskite \ce{CH_3NH_3PbI_3} was found to have promising photovoltaic performance~\cite{kojima2009organometal}, a surge of effort followed to investigate other perovskites in the hope of finding even better materials, for example by chemical tuning of the A-site cation and the X-site anion\cite{jung2019efficient, menendez2022mixed}. This has been a fruitful endeavour, with solid-solution perovskites reaching efficiencies of almost 27\%~\cite{jiang2025solar}. The same approach has also been applied to the search for other thin-film PV technologies and has led to a range of kesterite, Cu(In,Ga)Se and general chalcogenide materials~\cite{crovetto2020assessing, dimitrievska2026lessons, ma2025mapping, mauritz_precursor_2025, sopiha2022chalcogenide, blakesley2024roadmap, kayastha2025diverse}.

However, finding ``similar'' materials is an ill-defined task, dependent on how we represent the systems of interest. An early attempt at ``materials cartography'' constructed fragment-based features and measured distances to build connectivity graphs~\cite{isayev2015materials}. Compositional similarity can be measured through various means such as a distance (e.g. Euclidean) between vectors which represent the constituent elements. These vectors are referred to as embeddings, and the embedding for a compound could be, for example, a weighted sum of atomic embeddings\cite{park_mapping_2025, onwuli_ionic_2024}. Embeddings can be obtained in several ways. One common approach is to assign values to the vector based on the properties of the associated atoms, e.g. atomic number, electronegativity etc.\cite{ward_general-purpose_2016}.  Embeddings can also be learned; prominent examples include learning from scientific literature\cite{tshitoyan_unsupervised_2019}, learning from databases of chemical structures\cite{antunes_distributed_2022} and extracting them from deep learning models, such as transformers~\cite{wang_compositionally_2021} and graph neural networks~\cite{hagemann_transport_2025}.

Accounting for chemical and crystal structure in chemical representations is also critical. Methods such as many-body tensor representations (MBTR)\cite{huo2022unified} and Coulomb matrices build up a global representation of a structure by featurising the geometric relations between atoms\cite{rupp2012fast}. The smooth overlap of atomic potentials (SOAP) method builds representations of local structure, based on local atomic density\cite{de2016comparing}. 

Accounting for global structure and composition at the same time, and defining meaningful metrics for comparison is an open challenge. A recent pre-print from Negishi et al~\cite{negishi_continuous_2025} proposed a continuous metric to measure material dissimilarity, though this was intended for evaluating the uniqueness and novelty of materials predicted by generative models. Other methods have attempted to use embeddings generated from graph neural networks to capture structural similarity between materials\cite{xie2021crystal}. There are also strictly geometrical matching approaches such as the \texttt{StructureMatcher} method in the widely used \texttt{pymatgen} package\cite{ong2013python}. While this approach is useful for capturing exact matches, it does not provide a measure of similarity of systems that do not match exactly.  

Optimal Transport is a mathematical theory and framework for finding the most efficient way to transform one distribution (like a pile of dirt) into another (like a hole to fill), minimizing the total cost of movement. It provides powerful tools for comparing probability distributions, measuring distances between complex objects (like images or data sets), and enabling efficient data matching, with applications spanning computer vision, machine learning (GANs, diffusion models), fluid mechanics, economics, cell dynamics, and physics ~\cite{bunne_optimal_2024,levy_partial_2022}. Optimal transport-based approaches have been suggested for chemical use, first with the Earth Mover's distance from Hargreaves et al.~\cite{hargreaves_earth_2020}, although this compared materials using composition only (the Wasserstein, rather than Fused Gromov-Wasserstein, distance), and more recently an approach from Hagemann et al.~\cite{hagemann_transport_2025}, which used the Wasserstein distance to compare distributions of materials rather than the materials themselves.

In this work, we use the Fused Gromov-Wasserstein (FGW) distance from Vayer et al.~\cite{vayer_optimal_2019}. FGW uses optimal transport theory to find a mapping between the atoms and pairs of atoms (or nodes in their graphical representation) in two materials, and calculates a continuous distance metric which considers structural and chemical difference. We then correlate this distance with distances in the metric space of the spectroscopic limited maximum efficiency (SLME), a common heuristic for the power conversion efficiency (PCE) of a solar device. After establishing the best hyperparameters of the FGW model for matching to known SLME values, we use this method to search for new PV candidates. We start from known high-SLME materials and search materials databases for candidate systems with a low FGW distance, but which have not previously been explored for PV applications. We then take some of these candidate materials forward and obtain their SLME explicitly using hybrid-functional density functional theory (DFT) calculations.

We demonstrate that FGW provides a relatively inexpensive, but highly expressive approach to capturing meaningful similarities between materials. This work establishes FGW as a complementary approach to materials discovery alongside established approaches such as high-throughput screening and generative modelling~\cite{walker2026carbon, park2025exploration, cha2025learning, qin2024inverse, bone2025discovery}.

\section*{Methods}
\subsection*{Fused Gromov-Wasserstein for Crystals}
The Fused Gromov-Wasserstein (FGW) distance provides a metric for the dissimilarity between two materials, tuned by the parameter $\alpha \in [0,1]$, where 0 is a composition-only distance (the Wasserstein or Earth Mover's Distance) and 1 is a structure-only distance (the Gromov-Wasserstein distance). 
The FGW distance is based on a mapping or transport matrix $\pi$, where the matrix is iteratively updated to find the mapping with the smallest cost, and thus the `optimal transport'. Changing the original notation a little for clarity, the FGW distance minimises the cost $E_q$:
\[
FGW_{q,\alpha}(\mu,\nu)=\min_{\pi\in\Pi(h,g)}{E_q\left(M_{AB},C_A,C_B,\pi\right)},
\]
where
\[
E_q\left(M_{AB},C_A,C_B,\pi\right)=\langle(1-\alpha)M_{AB}^q+\alpha L\left(C_A,C_B\right)^q\otimes \pi,\pi\rangle
\]
\[
=\sum_{i,j,k,l}(1-\alpha)d(a_i,b_j)^q+\alpha\left|C_A(i,k)-C_B(j,l)\right|^q\pi_{i,j}\pi_{k,l}.
\]
 $M_{AB}=\left(d\left(a_i,b_j\right)\right)_{i,j}$ is a matrix of distances $d$ between feature vectors $a_i$ and $b_j$ on materials $A$ and $B$; a natural choice for this would be the Euclidean distance. $C_A$ and $C_B$ are matrices encoding the structure of each material through pairwise distances or connectivity of its atoms; the simplest choice would be adjacency matrices:
 \[   
C(i,j) = 
     \begin{cases}
       \text{1},\: i \text{ and } j \text{ connected,}\\
       \text{0,} \text{ otherwise.}
    \end{cases}
\]
 $\alpha$ is the Gromov parameter discussed above, while $q$ is left as 1 as in the original implementation, which ensures that the FGW has metric properties over the space of structured data~\cite{vayer_optimal_2019}. \\

Materials must be represented as graphs to use the FGW method: here, we used the \verb|StructureGraph| class from \verb|Pymatgen|, with the \verb|CrystalNN| strategy (based on the Voronoi algorithm) to estimate the atomic connectivities. Several feature vector types were considered, representative of a range of approaches:
\begin{itemize}
    \item One-hot vectors, the simplest approach where all atoms are maximally different to one another
    \item Randomly generated distributed vectors, providing more expressivity than one-hot vectors but containing no additional chemical information
    \item Magpie~\cite{ward_general-purpose_2016} and Oliynik~\cite{oliynyk_high-throughput_2016}, based on empirical data
    \item SkipAtom~\cite{antunes_distributed_2022}, learned distributed embeddings based on atoms usually found together in inorganic materials databases
    \item CrystaLLM~\cite{antunes_crystal_2024}, derived from the attention head of an LLM-based generative model
\end{itemize}
We also considered a number of structure matrix $C$ construction approaches and feature space distances $d$. We give these in full detail in the Supplementary Information (SI).

\subsection*{Photovoltaic data}
For photovoltaic discovery campaigns, we need a proxy for photovoltaic efficiency that balances accuracy with feasibility of calculation. The spectroscopic limited maximum efficiency~\cite{yu_identification_2012} (SLME) is a useful heuristic, using the frequency-dependent absorption coefficient and the minimum direct, dipole-allowed band-gap to give a more realistic estimate than the detailed-balance limit based on fundamental band-gap alone. This metric considers material thickness and temperature, which we keep as \SI{500}{\nano\meter} and \SI{300}{\kelvin}, respectively. More accurate heuristics have been proposed, including the Blank selection metric~\cite{blank_selection_2017}, but at present no large databases of material structures labelled with these heuristics exist, so the SLME remains the best approach for photovoltaic screening. In this study we use a dataset of SLMEs from Fabini et al.~\cite{fabini_candidate_2019}, which uses the $\Delta$-sol correction~\cite{chan_efficient_2010} to the generalised gradient approximation~\cite{perdew_generalized_1996} (GGA) for improved band-gap characterisation to calculate optical absorption spectra and thus SLMEs. This dataset contains 695 materials: far too few to train a neural network from scratch. This exemplifies the utility of the minimally supervised approach presented here, where inductive bias reduces the need for many data examples.

\subsection*{FGW hyperparameter optimisation}
The hyperparameters of the FGW were tuned to correlate with differences in SLME. FGW distances between all $N$ (695) materials in the dataset were calculated for each combination of parameters, resulting in a $\mathbb{R}^{695 \times 695}$ matrix for each. Then a $\mathbb{R}^{695 \times 695}$
matrix of differences in SLME between the materials was constructed. We min-max normalise both distance matrices then treat them as probability distributions, measuring the success of the FGW distances to capture distances in SLME space using the binary cross entropy (BCE) loss. The BCE loss between the true values $\left\{y_i\right\}$ (the SLME differences) and the predicted values $\left\{p_i\right\}$ (the FGW distances) is defined as:
\[
\text{BCE} = -\frac{1}{N}\sum_{i=1}^N{\left[y_i\log{p_i}+\left(1-y_i\right)\log{\left(1-p_i\right)}\right]}.
\]
 BCE plots were used to establish the best parameter choices for the subsequent steps.\\
\subsection*{$k$-Nearest neighbours regression}
The approaches with the best parameters for each feature vector were compared to three baseline methods using $k$-nearest neighbour regression, where properties are predicted by averaging the values of the $k$-nearest neighbours as implemented in  \verb|scikit-learn|.  The simplest baseline used Magpie~\cite{ward_general-purpose_2016} embeddings to featurise materials based on composition alone as implemented in the \verb|ElementEmbeddings| package. Next, we used the Smooth Overlap of Atomic Positions (SOAP)~\cite{bartok_representing_2013} as a similarity kernel $K(\cdot)$, from which we can construct the kernel distance~\cite{phillips_gentle_2011} as
\[
d(A,B)=K(A,A)+K(B,B)-2K(A,B),
\]
for materials $A$ and $B$. We use a standard parameter choice~\cite{olsthoorn_band_2019} of $n=8,\: l=6,$ and $ r_c=5$ \AA. Finally, we use Euclidean distances of embeddings from the Multi-Atomic Cluster Expansion (MACE)~\cite{batatia_mace_2022,batatia_design_2025} message-passing neural network (the `medium' pre-trained model available via the \verb|MACE| package) as a stand-in for the state-of-the-art of graph-level embeddings, also with a cut-off radius of 5 \AA, though message-passing means there is an effective radius of 10 \AA. We use mean pooling for the graph-level embedding, since SLME is an intensive property; indeed, this marginally improved RMSE in the KNN regression (see SI for a version of Figure~\ref{bce_knn}b with this comparison). Strictly, SLME is extensive, as it depends on material thickness, but the intrinsic photovoltaic efficiency is not, and a fixed value of thickness is used for all our reference values.\\

\subsection*{Clustering}
Clusters of chemical-structural similarity were found using $k$-medoid clustering. Unsupervised clustering is often performed with $k$-means clustering, but this requires coordinates for each point; here we only have pairwise distances. This clustering was performed using the \verb|kmedoids| Python package~\cite{schubert_fast_2022}. The number of clusters was optimised using two approaches: the mean silhouette score and the Kneedle algorithm~\cite{satopaa_finding_2011}. The silhouette score is defined as:
\[
S_\text{silhouette}=\frac{b_i-a_i}{\max(a_i,b_i)},
\]
where $a_i$ is the mean intra-cluster distance to point $i$ and $b_i$ is the mean nearest cluster distance to point $i$. Values range between -1 and 1, with the latter indicating perfect clustering. Meanwhile, the Kneedle algorithm finds the point of maximum curvature of a monotonically decreasing \textit{elbow} or \textit{knee} plot (so-named for their convexity and concavity, respectively), as implemented in the \verb|Kneed| Python package.\\

\subsection*{Materials discovery}
Once these clusters were formed, we took the top-SLME (a `seed' material) in each cluster and calculated the `local enrichment' $E(k)$, measuring the difference between the average SLME ($\eta$) among the $k$-nearest neighbours and the mean SLME of the entire dataset, to capture the extent to which good PVs are surrounded by other good PVs in the FGW distance space. This is effectively $k$-nearest neighbour regression looking at one material at a time, more closely replicating the task required for the screening process. For a seed $i$ and dataset-average SLME $\bar{\eta}$, we have
\[
E(k)=\frac{1}{k}\sum_{j\in \text{nn}(i)}^k \eta_j-\bar{\eta}.
\]

This analysis was followed by a material discovery campaign in which we calculated the FGW distances between our top-of-cluster materials and the entire Materials Project and then investigated structures in the vicinity of our seed materials.\\

\subsection*{Electronic structure calculations}
To verify the photovoltaic performance of the candidate materials, hybrid density functional theory (DFT) calculations were performed using the projected augmented wave (PAW) method~\cite{kresse_norm-conserving_1994,kresse_ultrasoft_1999} within the Vienna \textit{ab initio} Simulation Package (VASP)~\cite{kresse_ab_1993,kresse_efficiency_1996,kresse_efficient_1996}, using the Heyd–Scuseria–Ernzerhof functional HSE06~\cite{heyd_hybrid_2003}. The hybrid optics workflows were produced using \verb|atomate2|~\cite{ganose_atomate2_2025} under the independent particle approximation (IPA), which calculates dielectric properties reasonably well at a fraction of the cost of the more rigorous random phase approximation (RPA)~\cite{yang_high-throughput_2022}. \textit{k}-point meshes were made twice as dense as those generated by \texttt{atomate2} in each reciprocal space direction to ensure the convergence of optical properties. The \verb|VASPKIT| package~\cite{wang_vaspkit_2021} was used to calculate SLMEs at \SI{300}{\kelvin} for a material with thickness $d=\SI{500}{\nano\metre}$ as commonly used\cite{yu_identification_2012,fabini_candidate_2019}.
\section*{Results and Discussion}
\begin{figure}[htbp!]
\centering
\scalebox{1}[1]{
\includegraphics[width=1\textwidth]{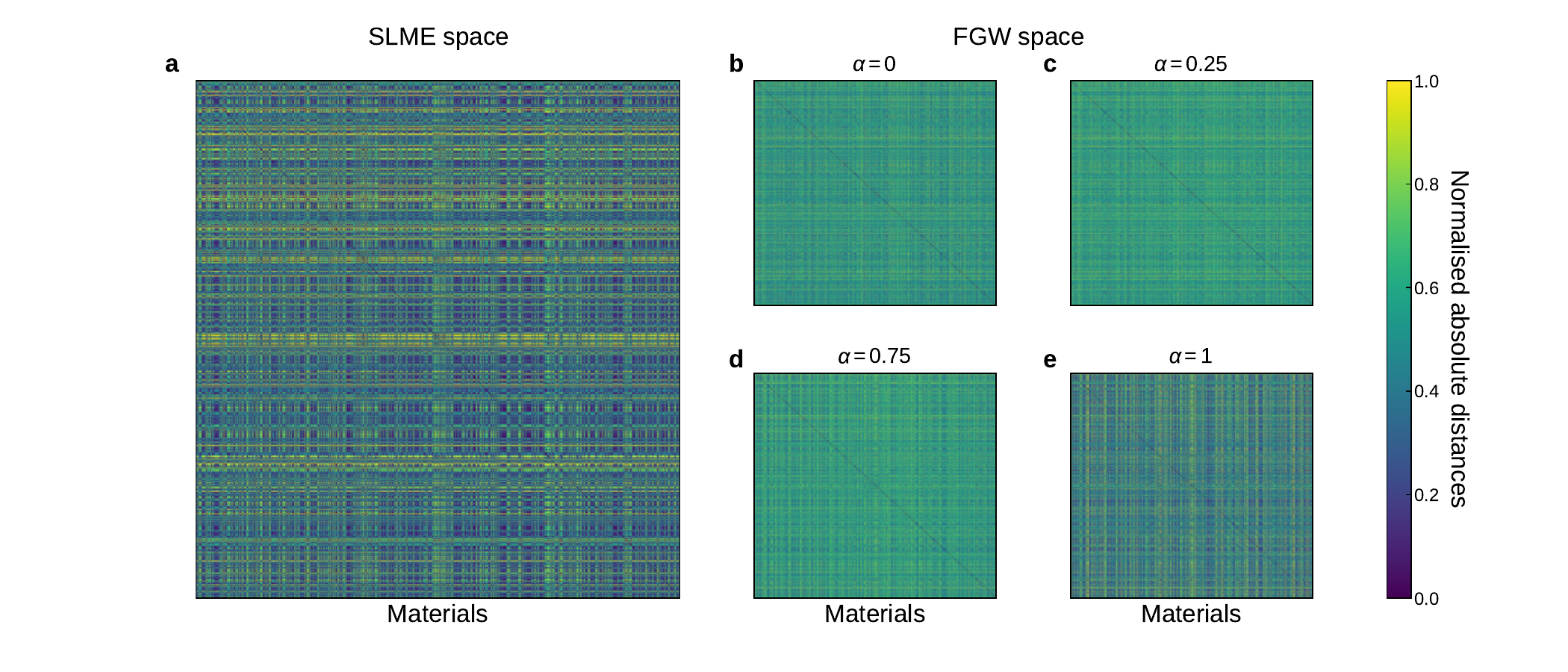}
}
\caption[Fig1]{Heat maps of normalised, absolute distances in \textbf{a}, SLME space and \textbf{b}--\textbf{e}, FGW space, where the Gromov parameter $\alpha$ is increased from 0 to 1 in \textbf{b} to \textbf{e} to include more structural importance to the FGW distance. The rows and columns on the $x$- and $y$-axes correspond to the materials in the dataset.}.
\label{heatmaps}
\end{figure}

Figure~\ref{heatmaps} shows heat maps of differences in SLME across the dataset and of FGW distances. We might expect that materials with small difference in chemical-structural features have similar optical properties. Thus, we can use the BCE loss, outlined in the Methods section, to measure how well the FGW distances calculated through a variety of parameter combinations capture the distribution of SLME differences in the dataset.\\

\subsection*{FGW hyperparameter optimisation}

\begin{figure}[htbp!]
\centering
\scalebox{1}[1]{
\includegraphics[width=1\textwidth]{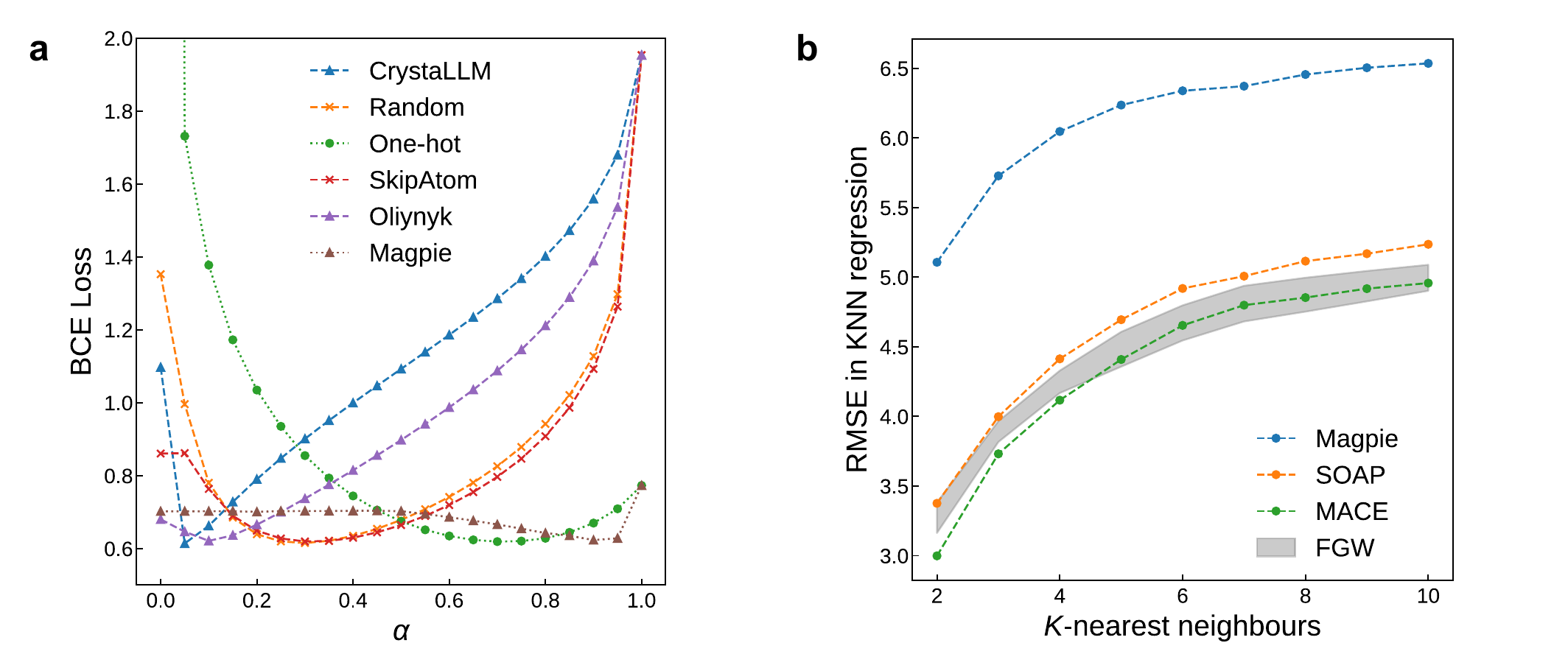}
}
\caption[Fig]{\textbf{a}, Plot of BCE loss against $\alpha$ for the parameter combination for each feature vector that resulted in the lowest minimum. Colours encode the feature vector, whilst dotted lines are the harmonic distance cost matrix and dashed lines use scaled atomic distances. Marker shapes indicate feature distance: circles for cosine distance, triangles for Euclidean distance, and crosses for Manhattan distance. Note the poor performances of the one-hot vector and the randomly generated distributed vector at $\alpha=0$ (composition-only). The one-hot curve extends to 18.1 here but was not shown for clarity. The legend lists the feature vectors from lowest minimum at the top (CrystaLLM) to highest at the bottom (Magpie), though there is very little difference between them (\SI{1.5}{\percent}). \textbf{b}, The same FGW methods at the optimal $\alpha$ values applied to a $k$-NN regression task, compared to three baselines for distance metrics: Euclidean distance between structure-level embeddings with Magpie \& MACE and the SOAP kernel distance.} 
\label{bce_knn}
\end{figure}

The results of hyperparameter tuning are summarised in Figure~\ref{bce_knn}a, where the BCE loss is plotted against the Gromov parameter $\alpha$ with the best parameter combination for each feature vector, measured by the lowest minimum BCE.  More detailed training results can be found in the SI or codebase. 
All plots show a minimum BCE at $\alpha\notin\{0,1\}$, indicating that the combination of compositional and structural cost is best for characterising photovoltaic activity, as we would expect. The one-hot and randomly generated distributed embeddings are particularly poor at $\alpha=0$, with the former reaching a loss of 18.1. This is to be expected from information-poor material representations although, interestingly, as $\alpha$ increases, the structural cost is sufficiently meaningful to reduce the loss to become comparable with the other embeddings. For all 6 feature vectors we use either the harmonic distance or atomic distance (scaled by the mean of the matrix) structure matrices. Both are more physically meaningful than the adjacency matrix. There is an even split among the feature distances considered: cosine, Euclidean, and Manhattan. All methods provide very similar minimum BCEs: $\alpha$ is likely the most important parameter to tune for any chemical application of the FGW distance.

\subsection*{$k$-nearest neighbour regression}
Taking the 6 FGW methods shown in Figure~\ref{bce_knn}a, we use \textit{k}-nearest neighbours regression to predict SLMEs of materials based on their $k$-nearest neighbours. The root-mean square error in the predictions is averaged across the dataset and plotted in Figure~\ref{bce_knn}b, with the grey area showing the upper and lower limits of the FGW methods. Also plotted are a number of alternative baseline approaches, outlined in the Methods section. The composition-only Magpie approach is significantly poorer than the rest at all $k$ values, reaffirming the importance of structure alongside composition for PV similarity learning, as also seen by the high losses at $\alpha=0$ in Figure~\ref{bce_knn}a. SOAP is a significant improvement with a nearly 50\% reduction in RMSE, due to the similarity kernel encoding structural information in the kernel distance between materials. The message-passing MACE provides an even better embedding, though the difference is much smaller than from Magpie to SOAP. The FGW methods are very similar to MACE. This highlights the sophistication of the FGW model: even one-hot vectors with a harmonic distance simply derived from the Voronoi connectivity, `trained' by varying $\alpha$ on 700 materials achieve comparable peformance to an embedding from an equivariant GNN pre-trained on $10^6$ bulk crystals. 

\subsection*{Clustering}

\begin{figure}[htbp!]
\centering
\scalebox{1}[1]{
\includegraphics[width=1\textwidth]{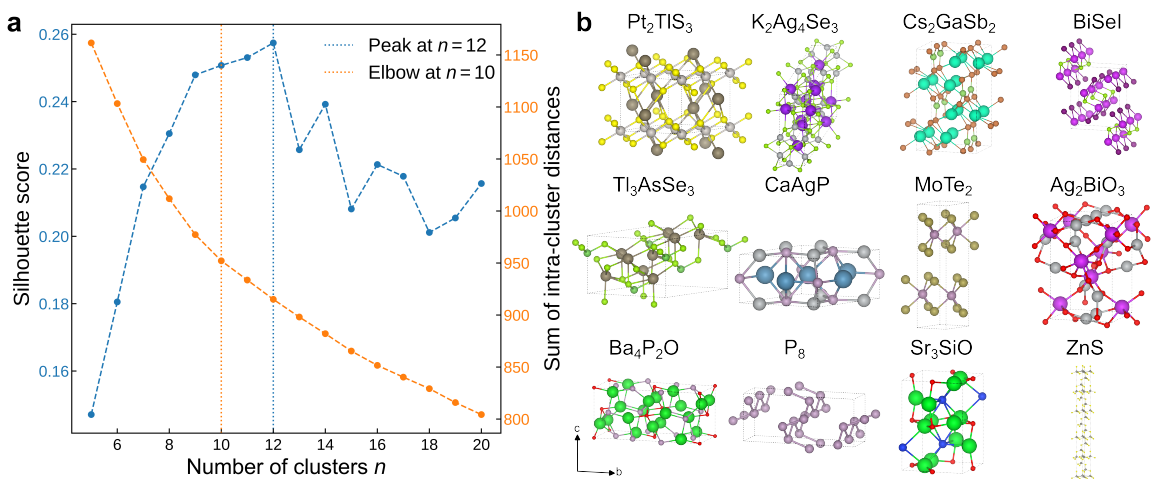}
}
\caption[Fig5]{\textbf{a}, Plots to identify the optimal clustering number through an elbow plot of the total intra-clustering distance and of the silhouette score, both against the cluster number, $n$. The optimal $n$ is found through the Kneedle algorithm for the elbow plot and by finding the peak of the silhouette score. Note the small discrepancy between the two approaches. \textbf{b}, Visualisations of the `seed' materials: the highest PV efficiency material from each cluster. Crystal axes $b$ and $c$ are shown for reference: $a$ varies based on bond angles.}
\label{cluster_plots}
\end{figure}
To ensure that the materials discovery campaign starts from a diverse set of seed materials, we first partition the existing data. The dataset was clustered using $k$-medoid sampling, as outlined in the Methods section, using the CrystaLLM embeddings and the associated parameters that gave the lowest BCE loss. Figure~\ref{cluster_plots}a contains plots to determine the optimal cluster number for the dataset using the maximum curvature of an \textit{elbow} plot and the peak of the silhouette score. The two approaches are in reasonable agreement, suggesting $10$ and $12$ respectively, and both $n$ values would give high silhouette scores and high curvature. $n=12$ was chosen because there is a clearer decision boundary for the silhouette approach. 

The material with the highest SLME in each cluster is given in Table~\ref{seeds_table} and displayed in Figure~\ref{cluster_plots}b. The seed materials in this table span a range of chemistries, including oxides, chalcogenides, pnictides, and mixed oxi-halides; they also span a range of space groups. This diversity of composition and structure gives an indication of the success of the FGW-based clustering in sampling a wide range of chemical space. All of these materials have SLME values above \SI{30}{\percent} - very high efficiencies - apart from the final material with an SLME of \SI{0.5}{\percent}. This suggests that the clustering has identified a group of materials with relatively similar chemical-structural embeddings that also all have extremely low SLMEs, further demonstrating the correlation between distances in FGW space and SLME space. 

\begin{table}[]
\centering
\begin{tabular}{@{}llll@{}}
\toprule
\textbf{MPID}      & \textbf{Composition} & \textbf{Space group} & \textbf{SLME} / \si{\percent} \\
\midrule
mp-9272   &      \ce{Pt_2TlS_3}       &  $P\bar{3}m1$           & 32.7      \\
mp-573891 & \ce{K_2Ag_4Se_3}         &  $C2/m$           & 32.3      \\
mp-29372  & \ce{Cs_2GaSb_2}            &  $Pnma$           & 32.2      \\
mp-23020  & \ce{BiSeI}            &  $Pnma$           & 32.1      \\
mp-7684   &  \ce{Tl_3AsSe_3}       &  $R3m$           & 31.3      \\
mp-12277  &  \ce{CaAgP}           &  $P\bar{6}2m$           & 31.2      \\
mp-602    & \ce{MoTe_2}         &    $P6_3/mcm$         & 31.1      \\
mp-23558  & \ce{Ag_2BiO_3}          &  $Pnn2$           & 30.8      \\
mp-28164  & \ce{Ba_4P_2O}        & $Cmce$            & 30.8      \\
mp-157    &   \ce{P_8}          &  $Cmce$           & 30.3      \\
mp-30949  &  \ce{Sr_3SiO}     &   $Pnma$          & 30.0        \\
mp-556775 &    ZnS         & $P3m1$            & 0.50
\\
\bottomrule
\end{tabular}
\caption[Tab1]{Table summarising seeds used for materials discovery campaign, found by taking the highest SLME material of each cluster. The final material belonged to a cluster of very low SLMEs and so was not used as a seed.}
\label{seeds_table}
\end{table}

\begin{figure}[htbp!]
\centering
\scalebox{1}[1]{
\includegraphics[width=1\textwidth]{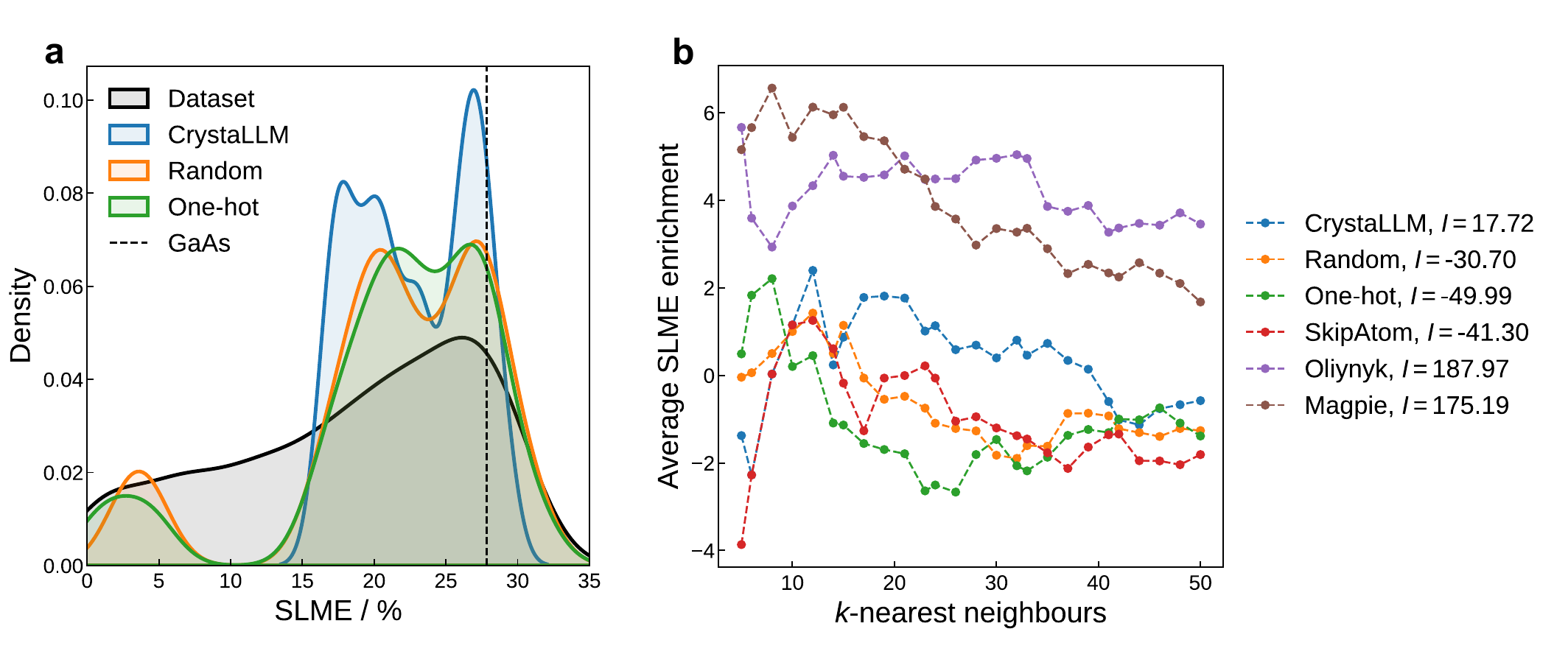}
}
\caption[Fig]{\textbf{a}, KDE plots showing the distribution of SLMEs for the entire dataset compared to the 20 nearest materials to high-PCE GaAs, using the best three FGW distance methods by BCE loss in Figure~\ref{bce_knn}a. The small unphysical densities at high and low SLMEs are artefacts of the kernel smoothing. \textbf{b}, `local enrichment' of the $k$-nearest neighbours of \ce{Pt_2TlS_3}, the seed with the highest SLME, in FGW space.}
\label{kde_enrichment}
\end{figure}

Figure~\ref{kde_enrichment}a demonstrates the effect of enrichment in the region close to a known high-performance PV absorber, namely GaAs. We plot kernel density estimates (KDE) of the distributions of materials close to GaAs using FGW with three different feature vector types. It is clear from the plots that the materials selected by FGW are significantly closer to the SLME of GaAs than the distribution of the underlying dataset. \\
Figure~\ref{kde_enrichment}b shows the enrichment for the highest-SLME seed for each of the top feature vector/parameter combinations: the difference between the average among the $k$-nearest neighbours and the total dataset average, with the integrals up to the first 50 neighbours given in the legend. A positive integral therefore indicates that materials in the vicinity of the seed (in FGW space) are generally better PVs than the dataset average, and a negative integral the opposite. There is a surprising amount of variation between embeddings, with a 50:50 split between positive and negative integrals, though there is skew towards higher values. For the remaining seeds there is generally closer agreement between feature vectors, as shown in the SI.

Interestingly, the two highest enrichment integrals belong to the Oliynik and Magpie embeddings, both hand-crafted from empirical data. This may be due to the fact that the learned embeddings already have some implicit contribution from crystal structure (as a result of how they are learned) and therefore do not work as well when combined with the explicit representation of structure in the Gromov distance. 
Based on these demonstrations of the capability of FGW distances to identify promising target materials, we now take the method forward for application in a materials discovery campaign. We use the Oliynik embeddings and parameters that gave the lowest BCE for them: Euclidean distance for $d$, scaled atomic distances for $C$, and $\alpha=0.1$.

\begin{table}[]
\centering
\begin{tabular}{@{}lllllll@{}}
\toprule
\textbf{Seed} & \textbf{MPID}& \textbf{Composition} & \textbf{$E_g$ / eV} & $E_g^\text{DA}$ \textbf{/ eV} &  \textbf{SLME} / \si{\percent} \\

\midrule
\ce{Ag_2BiO_3}   & mp-676840  & \ce{Ag_{13}(PbO_3)_6} &  0 & 0              &     N/A     \\
\ce{TlPt_2S_3}   & mp-28805   & \ce{Tl_2Pt_5S_6}      & 1.54   &   1.82           &   24.6      \\
\ce{Ag_2BiO_3}   & mp-28996   & \ce{Ag_4Bi_2O_5}      &   2.78   & 2.87            &   5.2       \\
\ce{Tl_3AsSe_3}  & mp-1196396 & \ce{Tl_4GeSe_4}      &    1.36   &  1.36         &   22.2       \\
\ce{MoTe_2}     & mp-267     & \ce{Te_2Ru}         &    1.09     & 1.71         &  26.4        \\
\ce{MoTe_2}     & mp-11675   & \ce{NbTe_2}         &  0          & 0      & N/A         \\
\ce{Tl_3AsSe_3}  & mp-28334   & \ce{Tl_4SiSe_4}      &  1.67      & 1.92          &  21.2        \\
\ce{TlPt_2S_3}   & mp-288     & \ce{PtS}    & 1.70       &  1.80   & 15.5         \\
\ce{Cs_2GaSb_2}  & mp-628742  & \ce{Cs_5Sb_8} & 1.16        &   1.17               &   31.3       \\
\ce{P_8}         & mp-569522  & \ce{MnP_4}*        & 0.97        & 1.03     & 24.7    \\
\ce{K_2Ag_4Se_3}  & mp-10480   & \ce{CsAg_5Se_3}*     & 1.30     & 1.38       & 26.4    \\
\ce{P_8}         & mp-16977   & \ce{Ti(MnP_6)_2}     &  0 & 0              &   N/A       \\
\ce{K_2Ag_4Se_3}  & mp-28468   & \ce{KAg_5S_3}        &        1.08       & 1.08   & 12.2          \\
\ce{CaAgP}     & mp-11214   & \ce{CaAgSb}        &       0.35     & 0.76     &  20.4        \\
\ce{BiSeI}     & mp-23066   & \ce{Pb_5(SI_3)_2}     &   2.45      & 2.45         &     5.74     \\
\ce{CaAgP}     & mp-5615    & \ce{CaAgAs}*        & 1.28         & 1.31   & 29.5    \\
\ce{BiSeI}     & mp-1197111 & \ce{Bi_20(PtI_{12})_3} &  1.62 &  1.63 & 22.0       \\
\ce{Cs_2GaSb_2}  & mp-542625  & \ce{Sr_3GaSb_3}*      & 0.77     & 0.94       & 14.8  \\ 

\bottomrule
\end{tabular}
\caption{Table summarising the PV candidates found by screening the Materials Project database, validated by DFT calculations using the HSE06 functional. Asterisks denote materials already present in the original dataset. The materials are sorted by FGW distance to their nearest seed -- we might expect that the SLMEs should decrease down the table accordingly.}
\label{PVs_table}
\end{table}
\subsection*{Materials discovery campaign}
To demonstrate FGW in a materials discovery setting, we use calculated FGW distances between the seed materials identified in Table~\ref{seeds_table} and a wider database of materials, which have not necessarily previously been considered as PV absorbers. This tests the capability of the FGW method to generalise beyond the data that were used to fix the hyperparameters.

The 155k materials in the Materials Project (as of January 2026) were downloaded, and FGW distances to each seed were calculated. The nearest 100 materials were taken for each seed. Materials containing lanthanides, actinides, or radioactive elements were removed. The candidates were then filtered to only include 3D materials with (GGA) band-gaps between 0 and 2.5 eV, energies above hull of below 50 meV, and compositions different from the seeds. Where multiple polymorphs of a non-seed material were found, only that with the lowest energy above hull was kept. Some materials were also found via multiple seeds, further reducing the pool of candidates. This left 186 materials of an original 1100, largely due to the 3-dimensional restriction -- full statistics of exclusion reasons are included in the SI. 2-dimensional structures can be excellent photovoltaics, but would introduce complications in the DFT calculations carried out to validate these predictions. Finally, the two materials closest to each seed were taken, although only one material remained for some seeds and none for mp-28164 and mp-30949. Of the 18 candidates, four were already in the Fabini dataset, leaving 14 to be validated with DFT calculations.\\

Table~\ref{PVs_table} summarises the results of the screening. We include the Kohn-Sham band-gap and the minimum direct, dipole-allowed band-gap to indicate the `directness' of the band-gap, crucial in the SLME formulation. A large difference indicates an indirect band-gap, which is typically detrimental to PV performance due to poor absorption of radiation. All but three materials are semiconductors: \ce{Ag_2BiO_3}, \ce{NbTe_2}, and \ce{Ti(MnP_6)_2} were found to be metallic at the HSE06 level, with experimental reports of the latter being metallic~\cite{scholz_synthese_1986}. A further two materials have SLMEs below \SI{10}{\percent} due to having large band-gaps. Given that we used a screening filter of up to 2.5 eV at the GGA level, which typically underestimates band-gaps, only two of the 18 materials having such wide gaps is a good result. Only 1 material of the original 1100 was excluded purely on having too wide a band-gap, with a further 20 excluded for stability/dimensionality reasons as well (see Supporting Information for a detailed breakdown of the screening process). Thus, the FGW approach biased the discovery space towards appropriate band-gaps effectively. There are three modest absorbers with SLMEs around \SI{15}{\percent}.\\

The remaining 10 materials have SLMEs above \SI{20}{\percent} -- considered high by Yu and Zunger~\cite{yu_identification_2012}. Three of these were in the dataset, so we propose 7 promising new materials. The most exciting is \ce{Cs_5Sb_8} with an SLME above \SI{30}{\percent}. Neither SLME nor hybrid DFT perfectly characterise PV performance but these are still very promising results for novel photovoltaics. 
We see the influence of the compositional matching in the FGW distance as every candidate material contains at least one atom from its seed. Some materials are simple monoatomic substitutions (such as \ce{CaAgSb} from \ce{CaAgP}), but some are significantly different, such as \ce{Bi_{20}(PtI_{12})_3} from \ce{BiSeI}, which would not be suggested by traditional, simpler similarity approaches.\\

A brief literature search using the search terms \texttt{<chemistry> photovoltaic} and \texttt{<chemistry> solar cell}, where \texttt{<chemistry>} is the stoiciometric formula reveals no prior work on any of these materials as photovoltaic absorber materials. This highlights the ability of the FGW guided database search to surface materials with non-trivial similarities to known high-performer materials. This is further emphasised by the complex crystal chemistry of the highest performing new material \ce{Cs_5Sb_8} , which has a mean charge on the Sb ions of -0.62, resulting from a range of Cs coordination environments and the amphoteric nature of Sb, which commonly adopts oxidation states of -III and III/V. This binary material is stable with respect to competing phases according to the Materials Project thermodynamics.

\section*{Conclusions}
We present a low-cost, minimally trained alternative to data- and resource-hungry deep learning approaches to materials screening. The inductive bias built into the FGW method through chemical embeddings and structure matrices lead to high similarity learning accuracies compared, for instance, to graph-level embeddings from graph neural networks trained on 1.6 M structures. An extensive grid search was carried out to find optimal settings for PV discovery. This was only possible due to the low cost of FGW calculations: averaging around 100 calculations core$^{-1}$ s$^{-1}$. Our results suggest that most of these choices make little difference and would not need to be explored so extensively in future study; the only parameter that requires some tuning is the Gromov parameter $\alpha$ which controls the relative importance of structure and composition. We then demonstrated the utility of the FGW method for materials discovery. Starting from seed structures, we search the Materials Project database for similar materials that have not previously been explored for PV application. We propose seven materials with high SLME, calculated by hybrid DFT, which to our knowledge have not received significant attention for PV application. This demonstrates the practical utility of FGW and provides a blueprint for using it in future materials discovery tasks.

\section*{Data availability}
The adaptation of the \verb|FGW| package from Vayer et al.~\cite{vayer_optimal_2019} we developed, \verb|ChemFGW|, is available at \url{https://github.com/mattheww98/ChemFGW}. Results and data analysis are available at \url{https://github.com/mattheww98/FGW_for_PVs}.


\footnotetext{\textit{$^{a}$Department of Chemistry, University College London, London WC1H 0AJ, United Kingdom. E-mail: matthew.walker.21@ucl.ac.uk}}
\footnotetext{\textit{$^{b}$~Department of Chemistry, University College London, United Kingdom. Email: k.t.butler@ucl.ac.uk}}
\footnotetext{\textit{$^{c}$~Department of Chemistry, University College London, United Kingdom.}}
\footnotetext{\textit{$^{d}$~Department of Engineering, University of Cambridge, United Kingdom.}}




\section*{Author Contributions}
Author contributions:
\begin{itemize}
    \item M. A. H. Walker: Data curation, Formal analysis, Investigation, Methodology, Visualisation, Writing - original draft, Writing - review \& editing
    \item Z. Zhou: Formal analysis, Investigation, Methodology
    \item J. Islam: Data curation, Formal analysis, Investigation, Methodology
    \item K. T. Butler: Conceptualisation, Funding acquisition, Supervision, Project administration, Writing - original draft, Writing - review \& editing
\end{itemize}
\section*{Conflicts of interest}
There are no conflicts to declare.
\section*{Acknowledgments}
We acknowledge support from EPSRC project EP/Y000552/1 and  EP/Y014405/1. Via our membership of the UK's HEC Materials Chemistry Consortium, which is funded by EPSRC (EP/X035859/1), this work used the ARCHER2 UK National Supercomputing Service \url{http://www.archer2.ac.uk}. The authors acknowledge the use of the UCL Myriad High Performance Computing Facility (Myriad@UCL), and associated support services, in the completion of this work. We acknowledge Young HPC access which is partially funded by EPSRC (EP/T022213/1, EP/W032260/1 and EP/P020194/1). ZZ was funded by an AIchemy Summer Studentship, funded by UKRI (EPSRC grant EP/Y028775/1 and EP/Y028759/1).




\renewcommand\refname{References}


\bibliography{references, references-ktb} 

\bibliographystyle{rsc} 

\end{document}